\documentstyle[amsmath,11pt,graphicx]{article} 
\newif\ifpdf\ifx\pdfoutput\undefined\pdffalse\else\pdfoutput=1\pdftrue\fi
\newcommand{\pdfgraphics}{\ifpdf\DeclareGraphicsExtensions{.pdf,.jpg}\else\fi}

\begin{document} 
\pdfgraphics
 
\title{Opening of DNA double strands by helicases.\\ 
Active
 versus passive opening} 

\author{M. D. Betterton$^{1,2}$ and Frank J\"ulicher$^{1,3}$ \\ 
$^1$ Institut Curie, Physico-chimie Curie, Paris\\
$^2$ Courant Institute, New York University, 251 Mercer
St., New York, NY 10012 \\
$^3$ Max-Planck Institute for physics of complex systems\\ N\"othnitzerstr. 38,
01187 Dresden, Germany}

\maketitle 
\abstract{Helicase opening of double-stranded nucleic acids may be 
``active'' (the helicase directly destabilizes the dsNA to promote
opening) or ``passive'' (the helicase binds ssNA available due to a
thermal fluctuation which opens part of the dsNA). We describe
helicase opening of dsNA, based on helicases which bind single NA
strands and move towards the double-stranded region, using a discrete
``hopping'' model.  The interaction between the helicase and the
junction where the double strand opens is characterized by an
interaction potential.  The form of the potential determines whether
the opening is active or passive. We calculate the rate of passive
opening for the helicase PcrA, and show that the rate increases when
the opening is active. Finally, we examine how to choose the
interaction potential to optimize the rate of strand separation.  One
important result is our finding that active opening can increase the
unwinding rate by 7 fold compared to passive opening.}

\section{Introduction} 
A helicase is a molecular-motor protein which opens double-stranded
nucleic acid (NA) molecules. The strand separation is fueled by
nucleoside triphosphate (NTP) hydrolysis, typically of ATP. Since the
discovery of the first helicase in 1976 \cite{mon76}, more than 60
distinct helicase proteins have been found \cite{loh96}, including 12
apparent helicases solely in the bacterium {\it E. coli}. Helicases
play a role in nearly every cellular process which involves NA,
including replication, transcription, translation, repair, and RNA
processing \cite{loh96}.  All helicases share the ability to move
along NA strands, and to couple that motion to strand separation. For
this reason, helicases are often referred to as DNA translocases
\cite{sin02}. Certain helicases, because of their role in repair of
DNA damage, are crucial to maintain a stable (undamaged) genome. For
example, helicase mutations play a role in the human diseases
Xeroderma Pigmentosum, Bloom syndrome and Werner syndrome, all of
which are associated with a predisposition to cancer
\cite{bra00}.

Given the ubiquity of helicases in cellular processes which access
nucleic acids and the large number of different types of helicase, it
is not surprising that a variety of different helicase structures
exist.  We note that the many different examples of helicases are not
all similar, and their unwinding mechanisms are not necessarily the
same. The earliest known helicases were ring-shaped
hexamers\cite{loh96}; other helicases have been shown to function as
dimers. In the last 5 years, the helicase superfamilies I and II (SF1
and SF2, based on sequence comparison) have been discovered; this work
led to a revision of the view that helicase function requires a
multimeric protein. For a review of the SF1 and SF2 helicases see
\cite{sin02}.
 
This paper was inspired by experiments on the SF1 and SF2
helicases. The monomeric proteins are known to bind to single-stranded
(ss) NA and translocate directionally along it. Upon reaching the
ss-double strand (ds) junction, the helicase continues and moves the
junction foward, creating additional ssNA ``track'' as it goes.  In
this paper, we describe the physical principles of this process and
show how to optimize performance of the enzyme.  The coupling between
the helicase translocation and NA opening is loosely classified in the
literature \cite{loh96} as ``passive'' or ``active.'' Passive
unwinding means that the helicase unwinds indirectly, by binding ssNA
available due to a fluctuation which opens part of the dsNA. We also
call this ``hard'' unwinding for reasons discussed below, and to avoid
confusion with other meanings of passive/active. In active unwinding
(which we refer to as ``soft'' unwinding) the helicase directly
destabilizes the dsNA, presumably by binding to the dsNA and changing
the free energy of the ds state.
 
Experiments on the SF1 protein PcrA support a soft
mechanism. Velankar {\it et al.} \cite{vel99} solved crystal
structures of the enzyme bound to a forked (part ss and part ds) DNA
substrate.  In one of these structures, the protein binds the
non-hydrolyzable ATP analog ADPNP and represents the helicase prior to
ATP hydrolysis. A second structure contains a sulfur ion and
represents the conformation after ATP hydrolysis.  These structures
showed that the helicase binds ssDNA and tracks along the single
strand as it moves. However, in one crystal, a portion of the PcrA
also binds to the duplex DNA ahead of the ss-ds junction, causing a
distortion of the helical structure. The authors suggested that this
distortion promotes melting of the helix, thereby facilitating strand
separation. The crystal structures and footprinting studies
\cite{sou00} showed that the contact between the protein and the double 
helix occurs 4-5 base pairs (bp) from the junction and again 12-13 bp
from the junction.  Soultanas {\it et al.} mutated the residues of
PcrA which interact with the duplex portion of the DNA
\cite{sou00}. The mutant proteins unwound dsDNA 10-30 fold more
slowly than the wild-type protein, depending on which residue is
mutated. These experiments provide evidence that PcrA unwinds
softly and suggest that hard unwinding may be slower and
less effective than soft.
 
In this paper, we use a simple model to explore the theoretical 
consequences of soft and hard opening.  We suppose that the 
helicase and the NA ss-ds junction interact.  Physically, the 
interaction corresponds to an interaction potential which describes 
how the presence of the helicase modifies the free-energy change of 
the NA opening.  In our formulation, passive opening corresponds to a 
hard-wall potential: the helicase cannot advance beyond the ss-ds 
junction, but the presence of the helicase does not affect the 
energetics of dsNA opening.  Softer potentials correspond to active 
opening, because the helicase changes the local energetics at the 
junction. 
 
We study how NA strand separation depends on the interaction. In 
particular, we show that passive opening is {\em always} slower than a 
simple form of active opening. We calculate the rate of hard 
opening for the helicase PcrA, and discuss how the rate increases for 
soft opening. Finally, we examine how to choose the interaction 
potential to optimize the rate of strand separation.  One important 
result is our finding that soft opening can increase the unwinding 
rate by 7 fold compared to hard opening. This result is consistent 
with the mutation studies of PcrA discussed above. 
 
Although extensive biochemical and structural studies of helicases
have been performed, relatively little has been done to characterize
the function of helicases theoretically.  An important review of
helicase mechanism by von Hippel and Delagoutte \cite{von01}
synthesizes the thermodynamic and kinetic properties of helicases into
a general framework and emphasizes the importance of considering
helicase-NA binding, translocation, and strand-separation
independently. Doering {\it et al.} \cite{doe95} developed a
``flashing field'' model specific to hexameric ring
helicases. Recently Bhattacharjee and Seno \cite{bha02} considered the
coexistence of ss and dsDNA in the presence of a helicase. Chen {\it
et al.} \cite{che97} described a helicase as a biased random walk, and
considered how the density of histones affects the random
walk. However, we know of no theoretical work which addresses how the
interaction between helicase and NA affects the unwinding in general,
or the difference between active and passive unwinding in
particular.  Helicases offer a particularly appealing choice for a
simplified physical description because they are an important class of
enzyme which interacts with NA, yet they have many relatively simple
features. For example, many helicases have unwinding rates which are
independent of the NA sequence unwound\cite{loh96,dil00}. Thus, the
information content of the NA does not play an essential role in
helicase operation---a large simplification compared to RNA
polymerases, for example\cite{nud94,upt97}.  Monomeric helicases are a
relatively simple system that may serve as a building block for future
theoretical studies of helicases.  However, the principles discussed
here are quite general and could also apply to multimeric helicases.
 
\subsection{Helicase properties} 
 
Extensive experiments in the last thirty years have characterized
several important properties of helicases\cite{loh96}. A helicase may
bind a ss tail of NA adjoining a duplex, a forked NA containing both a
ds region and two ss arms, or the blunt end of dsNA. Most helicases
bind and translocate with polarity: they move---either on a single
strand, or on one of the two strands of a duplex---in a preferred
direction. A 3'-5' helicase moves toward the 5' end of the NA strand
to which it is bound.  The monomeric SF1 helicases we consider here
bind to a ssNA tail. Thus after binding, the helicase moves along the
ss tail to the ss-ds junction, then continue moving to unwind.

Once a helicase has bound to a strand and begins moving, it unwinds 
at a measurable average rate, ranging from a few (for PcrA) to 
thousands (for RecBCD) of bp/sec. The average 
number of base pairs unwound before the helicase falls off the NA is 
known as the processivity; measured values for different types 
of helicases vary from 40 to 30,000 bases unwound 
\cite{loh96}.  Both the rate and the processivity of a helicase are 
strongly affected by the presence of accessory proteins as well as
solution conditions. For example, {\it E. coli} single-strand binding
protein (SSB) decreases the average free energy cost of unwinding a
segment of dsDNA, and has also been observed to increase the
processivity of helicases.  High values of processivity, such as the
30,000 bp measured for RecBCD, are usually observed in the presence of
SSB\cite{loh96}. Hexameric ring helicases also tend to have higher
processivities than dimeric or monomeric helicases.  In the absence of
any accessory proteins, the monomeric helicase PcrA translocates on
ssDNA at a rate of about 80 bases/s \cite{dil02} and has a low
processivity of about 50 bp unwound \cite{sou99}.
 
A helicase is able to unwind dsNA because it uses the energy released
by nucleotide hydrolysis (usually by ATP) for its motion.  The average
number of bp unwound per ATP molecule hydrolyzed is called the
efficiency of unwinding (this quantity is often called the
``efficiency" of the helicase in the literature; we prefer to call it
the efficiency of unwinding to distinguish it from the efficiency of
energy transduction, which is related but not exactly the same).
Again, the presence of SSB changes the free energy balance and
increases helicase efficiency of unwinding.  Under {\it E. coli}
conditions at room temperature, ATP hydrolysis releases about 23 kT,
and the average free-energy change from melting one base pair is
approximately 1.7--2.5 kT \cite{loh96}. The theoretical limit on
efficiency of unwinding is thus 9--12 bp unwound per ATP
hydrolysis. Experimentally measured values are typically smaller,
usually 0.33--1 bp unwound/ATP hydrolyzed\cite{loh96}. For PcrA, the
measured efficiency of unwinding is approximately 1 bp unwound/ATP
\cite{dil00}. 
 
\begin{figure}[t] 
\centering
%\centerline{\epsfysize=4.0cm\epsfbox{hopdna}} 
\includegraphics[height=5cm]{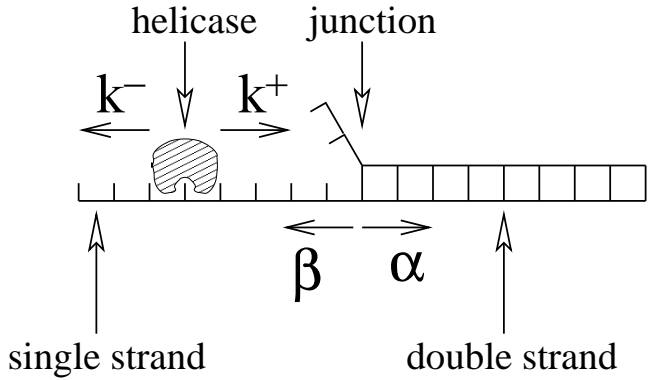}
\caption[] 
{Sketch of helicase on NA.} 
\label{hopdna} 
\end{figure}

\section{Hopping Model} 
\label{hop} 
 
We formulate a discrete model for helicase motion and strand
separation (Fig. \ref{hopdna}).  We consider two degrees of
freedom: the helicase position $n$ along the NA strand and the
position $m$ of the NA ss-ds junction. Since we are considering only
single-base hops, both $n$ and $m$ are integers. (In our units the
length of a ss base is 1.) The helicase hops one base forwards towards
the dsNA with rate $k^+$, and backwards with rate $k^-$. Since the
helicase hydrolyzes ATP, it is out of equilibrium\footnote{The rates
$k^+$, and $k^-$ can be derived from a more complicated model which
explicitly treats the nonequilibrium nature of the enzyme. A
forthcoming article will address this issue.} and has
$k^+>k^-$. Similarly, the NA opens by one base\footnote{We consider
only the breathing modes of the NA in which it opens or closes by one
base pair. Modes in which larger number of bases open are present, but
they are much less probable than the single-base opening or closing.}
with rate $\alpha$ and closes with rate $\beta$. Because the NA tends
to close (in the absence of melting agents \cite{von01}) we have
$\beta > \alpha$. Thus the motion of the helicase and the junction
tend to drive their positions closer together.

Our simple picture neglects several effects. We ignore binding and
unbinding of the helicase, and thus consider the motion of the
helicase along the NA strand only. The NA strand we treat as a rigid,
one-dimensional track (twist relaxation is fast compared to the
unwinding rate). We neglect effects of the NA sequence; each NA base
is identical.  We ignore the different biochemical states of
the helicase and describe it only by forward and backward rates.
 
For a helicase to unwind a NA duplex, it must interact with (and move)
the ss-ds junction. The form of this interaction determines whether
the unwinding is soft or hard.  We describe this interaction
by an potential $U(m-n)$ between the helicase and the NA ss-ds
junction which depends only on the difference between $m$ and $n$.
For large separations $m-n \gg 1$ we assume that the junction and the
helicase have no effect on each other, so the coupling potential tends
to zero. However, when $m-n$ is small the coupling potential $U$
changes both the helicase motion and the junction motion: $U>0$
inhibits the forward motion of the helicase and increases the relative
opening rate of the junction.
 
We use detailed balance (the law of mass action) to write how this
coupling potential changes the rates. If the NA closes (so $m \to
m-1$), the change in interaction energy is $ U(m-n)-U(m-n-1)$. The
opening and closing of the NA happens without an external energy
source, so detailed balance applies. The ratio of the opening and
closing rates is
\begin{equation} 
\frac{\beta_{m}}{\alpha_{m-1}} = \frac{\beta}{\alpha} 
	e^{-\frac{U(m-n-1)-U(m-n)}{kT}}, 
\label{posdependence} 
\end{equation} 
where $\alpha$ and $\beta$ are the sequence-averaged rates when the 
helicase and ss-ds junction are far apart. (See Section \ref{disc} for 
a discussion of the values of the different rates.)

We now apply the same argument to the helicase motion.  The change in 
energy $U(m-n-1)-U(m-n)$ represents the effective force acting on the 
helicase. We describe the force dependence of helicase motion by 
writing the ratio of forward and backward rates as 
\begin{equation} 
\frac{k^+_{n}}{k^-_{n+1}} = \frac{k^+}{k^-} 
	e^{-\frac{U(m-n-1)-U(m-n)}{kT}} 
\label{posdepk} 
\end{equation} 
Eq. (\ref{posdepk}) is not valid in general, because the helicase 
is driven by ATP hydrolysis and detailed balance cannot be 
used. However, it holds for certain cases in a model for motion 
generation by ATP hydrolysis\footnote{ A
forthcoming article will address this issue.} 
and is very useful for a discussion of the principles.

\begin{figure}[t] 
\centering
\includegraphics[height=5cm]{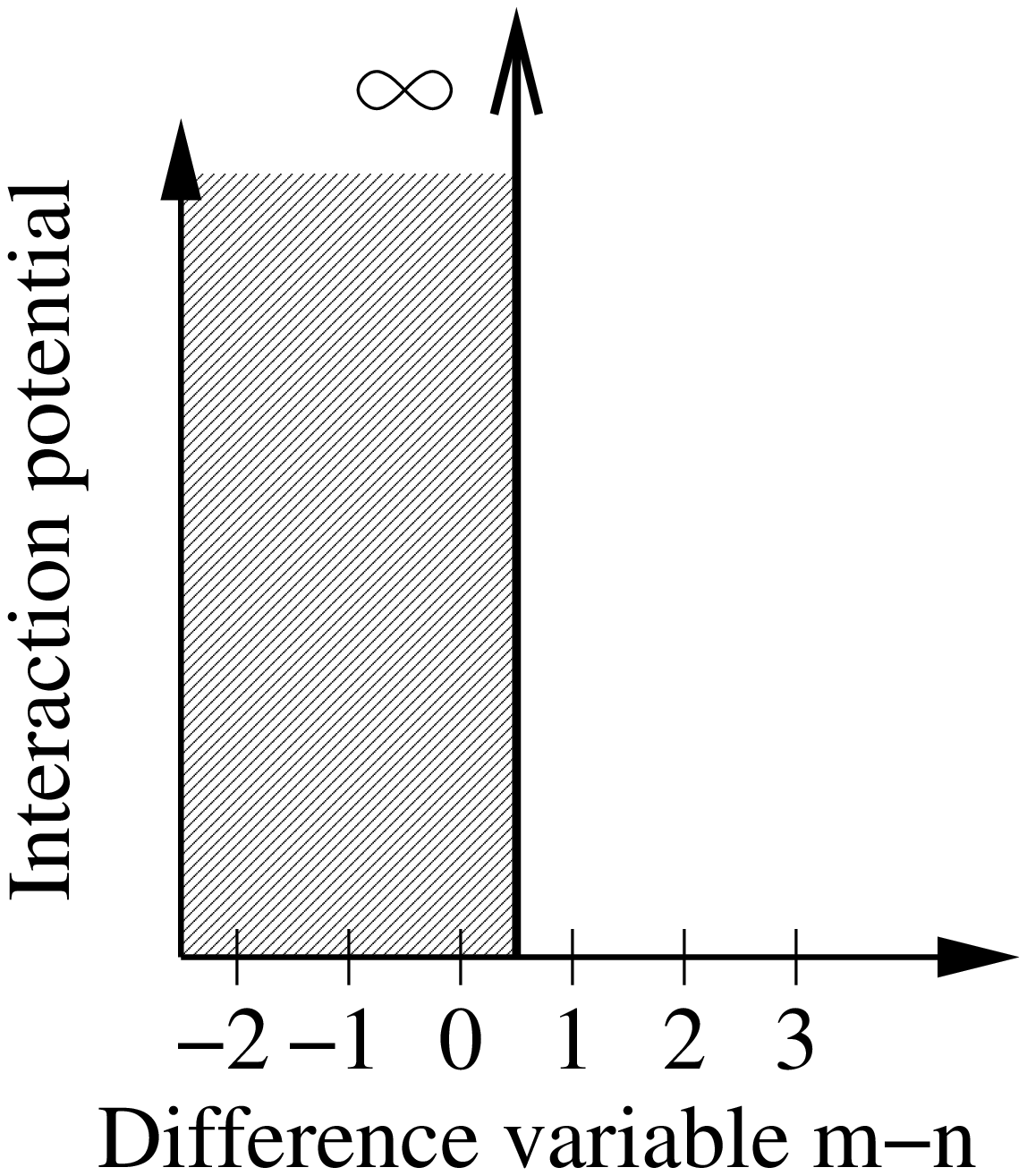}\includegraphics[height=5cm]{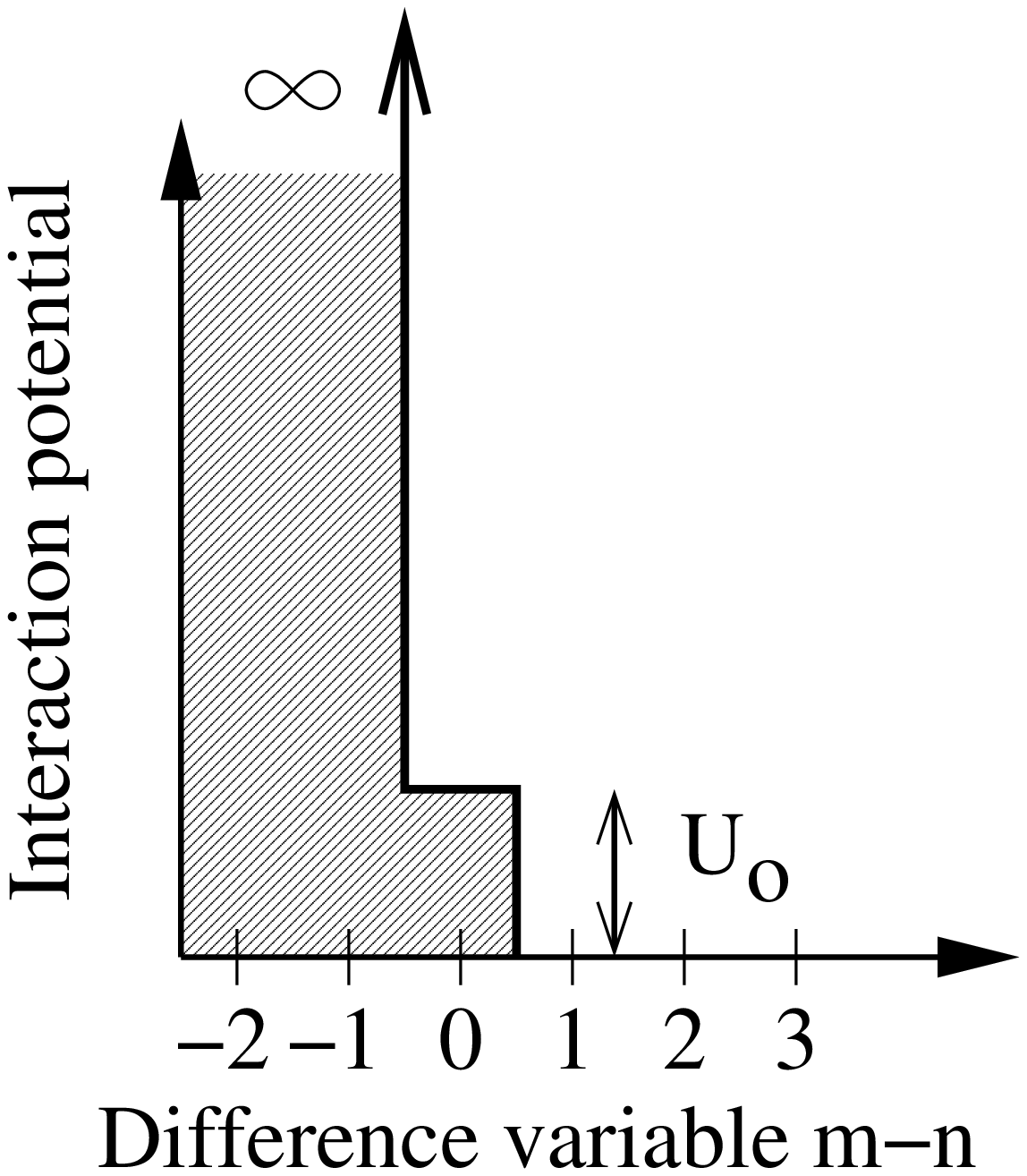}\includegraphics[height=5cm]{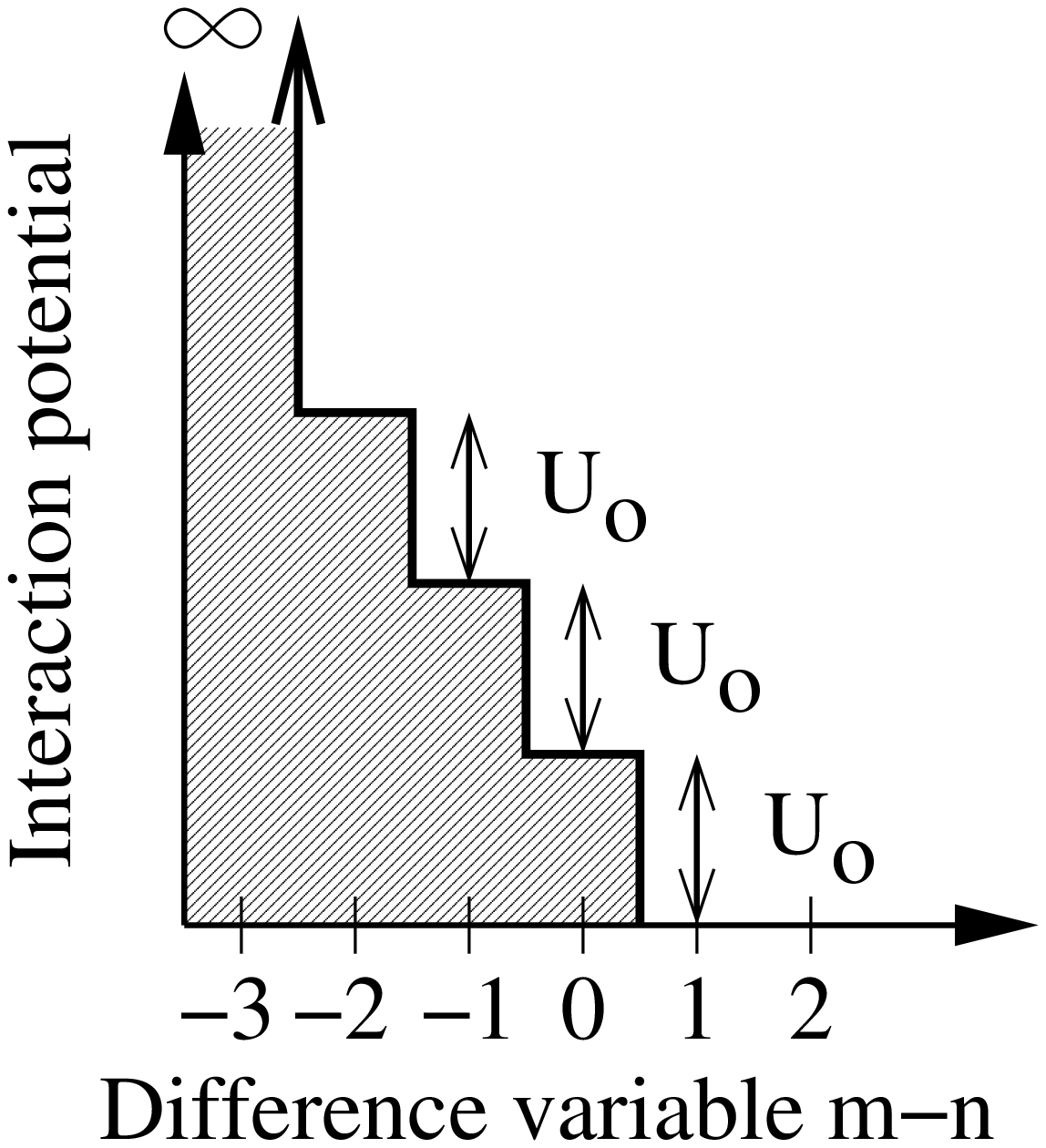}
\caption[] 
{Schematic interaction potentials between the helicase and ss-ds
junction (a) Left, hard wall potential corresponding to passive 
opening. (b) Middle, a potential with a step, which corresponds to 
active opening. (c) Right, a three-step potential.} 
\label{enplot} 
\end{figure}

The essence of passive opening is to suppose that the helicase 
provides a steric inhibition of NA closing when it is near the 
junction. If the junction is at position $m=n+1$, then the helicase 
covers the ss base adjacent to the junction and prevents base-pairing 
with the complementary NA strand. Thus hard opening requires 
$\beta_{m}$ to equal 0 when $m=n+1$. By  equation 
(\ref{posdependence}), we see that  $\beta_{m}=0$ 
if  $U(0)$ is infinite. This is the 
hard-wall potential sketched in Fig. \ref{enplot}a. 
 
The effect of this potential on the helicase motion makes intuitive
sense, as seen in Eq. (\ref{posdepk}). If $U(0)$ is infinite then
$k^+_{n}=0$ when $m =n+ 1 $. In other words, the helicase is unable to
advance into the double-stranded NA region and must wait until a
thermal fluctuation opens the NA before advancing. This example shows
how we describe hard opening: when the helicase and dsNA cannot
overlap, the interaction energy $U$ is infinite for $n \ge m$.
 
The interaction potential $U(m-n)$ can be ``softer'' than the hard
wall form.  Such a potential (as in Fig. \ref{enplot}b) leads to
active opening, because the interaction with the helicase increases
the opening rate of the dsNA. Recall Eq. (\ref{posdependence}) for the
position dependence of the rates. Whenever the NA closing increases
the interaction energy ($U(m-n-1)>U(m-n)$), the ratio of NA closing to
opening rate decreases, relative to $\beta/\alpha$. Thus the
interaction with the helicase can increase the rate of opening or
decrease the rate of closing (or both). The crystal structure of PcrA
discussed above, which shows that the enzyme can bind to the dsDNA and
distort it suggests active opening. In addition, point mutations of
the region of PcrA which binds the duplex DNA decrease the unwinding
rate \cite{sou00}. In our picture, the mutations could alter the
potential so it approaches a hard-wall form (which corresponds to
hard opening).
  
For a real helicase such as PcrA we do not know the exact  
form of the potential. We explore different forms of $U(m-n)$ subject  
to the requirements (i) $U \to 0$ for $m \gg n$ (when the helicase and  
the junction are far apart, there is no interaction) and (ii) $U  
\to \infty$ for $n \gg m$. The second requirement means that the  
helicase cannot bind solely to dsNA and always retains some
``footing'' on the ssNA. An increase in the energy for $n>m$ is
necessary to confine the helicase near the ss-ds junction and prevent
it from wandering off into the ds region.
  
In the analysis of this model, we address several questions. How does
changing the potential affect the unwinding rate? We showed above that
an increasing interaction potential has two effects: it leads to
relatively faster NA opening and relatively slower helicase
forward motion. We wish to understand how this trade-off can be
resolved. In particular, is there a choice of interaction potential
which leads to the fastest possible rate of opening? We address these
questions below for certain simplified forms of the potential $U$.
  
\subsection{Solving for the opening rates}  
  
We wish to determine the rate of dsNA unwinding for the hopping model,
given the rates $k^+$, $k^-$, $\alpha$, $\beta$, and the potential
$U(m-n)$. We consider the probability $P(n,m,t)$ that the helicase is
at base $n$ and the ss-ds junction at base  $m$ at time $t$. We
give the details of this calculation in App.
\ref{hopderive} and outline the main features here. Our approach is  
similar to the polymerization ratchet of Peskin, Odell, and 
Oster, who examined how a polymerizing filament is able to produce a 
force \cite{pes93}. In their work, the two fluctuating degrees of 
freedom are the tip of a filament and a wall, which are 
analogous to our helicase and ss-ds junction.  However, they considered 
a hard-wall interaction; thus our work may be considered a 
generalization of their approach. The probability distribution 
satisfies the 
Eq. (\ref{basicprob}), which describes the changes to 
$P(n,m,t)$ due to opening/closing of the NA (which changes $m$) and 
helicase hopping (which changes $n$). 

We rewrite this equation using difference and midpoint variables 
$j=m-n$ and $l=m+n$ (Eq. (\ref{newprob})). This allows an 
important simplification since the coefficients in Eq. (\ref{newprob})
are independent of $l$. By summing Eq. (\ref{newprob}) over $l$,
we obtain Eq. (\ref{jprob})  for the difference-variable
distribution ${\cal P}_j=\sum_l P(j,l)$. This equation describes
the relaxation of 
the difference-variable distribution to a stationary
one, which corresponds to a constant current in the difference
variable $j$.
The potential diverges as $j \to -\infty$, which implies that the helicase
cannot pass the junction. Therefore,
the $j$-current must vanish, which 
leads to a simple recursion relation for ${\cal P}_j$ that depends on the
interaction potential.

The strand separation velocity can be calculated assuming that the
distribution ${\cal P}_j$ is stationary.  The distribution ${\cal
P}_j$ reaches its steady state after a relaxation time related to the
rates $\alpha$, $\beta$, $k^+$, and $k^-$.  Summing
Eq. (\ref{newprob}) over $j$ leads to a discrete drift-diffusion
equation for $\Pi_l=\sum_j P(j,l)$ (Eq. (\ref{lprob})). The
corresponding current in $l$ does not vanish and determines the mean
strand separation velocity
\begin{equation}  
v = \frac{1}{2} \sum_j (k^+_{j} +\alpha_{j}-k^-_{j} -\beta_{j}){\cal P}_j.  
\label{vel}  
\end{equation}  
The expression for $v$ has a simple 
physical interpretation.  The quantity in parentheses is the unwinding 
rate at separation $j$---either a forward hop of the helicase 
($k^+_j$) or NA opening ($\alpha_j$) move the 
helicase/junction complex forward. Similarly, a backward hope of the 
helicase ($k^-_j$) or  NA closing ($\beta_j$) move the 
complex backwards. We then take the unwinding velocity at separation 
$j$ and multiply it by the probability ${\cal P}_j$ of finding the 
complex at separation $j$. Repeating this addition for all possible 
$j$, we arrive at the total unwinding rate.  Thus we see that, under 
our assumptions, solving Eq. (\ref{nocurr}) for the stationary 
probability distribution ${\cal P}_j$ is suffient to calculate the mean 
helicase opening velocity.
The diffusive part of the equation for $\Pi_l$ allows us 
to determine the effective diffusion coefficient which  
characterizes velocity fluctuations  
\begin{equation}  
D=\frac{1}{4} \sum_j (k^+_{j} +\alpha_{j}+k^-_{j} +\beta_{j}){\cal P}_j  
\end{equation}  
  
\section{Hard versus soft opening}  
 
Consider  the hard wall model for passive opening, with the 
wall at $j=0$. As described above, this means $k^+$, $k^-$, $\alpha$ 
and $\beta$ are constant in the region $j>0$ except at $j=1$, where 
$k^+_{1}=\beta_{1} = 0$. Since the rates are constant, we have 
\begin{equation} {\cal P}_j = A  
\left(\frac{\alpha+k^-}{\beta+k^+}\right)^j = A\ c^j,  
\end{equation}  
where we have defined $c=\frac{\alpha+k^-}{\beta+k^+}$ and the
constant $A$ is determined from normalization:
$A=(\beta+k^--\alpha-k^-)/(\alpha +k^+)$.  Evaluating Eq.(\ref{vel})
for the helicase opening velocity, we find
\begin{equation}  
v_{HW}=\frac{\alpha  k^+ - \beta k^-}{\beta+k^+}  
\label{vhardwall}  
\end{equation}  
The opening velocity is positive whenever $k^+/k^- > \beta/\alpha$, 
that is, the free energy change which drives DNA closing must be 
smaller than the effective free energy change of the helicase.  This 
requirement intuitively makes sense: the helicase must have a 
sufficiently high forward rate to overcome the energetically favorable 
DNA closing. The sequence-averaged value of $\beta/\alpha$ for DNA is 
7 (corresponding to an average free energy change per base closed of 2 
kT). Thus, even by this simple calculation we see that for passive 
opening to be possible we must have $k^+ > 7 k^-$. The backward 
rate $k^-$ must be considerably smaller than the forward rate, as 
observed in experiments\cite{dil00,dil02}.

\subsection{Soft opening: the staircase}  
 
\begin{figure}[t]  
\centering
\includegraphics[height=5cm]{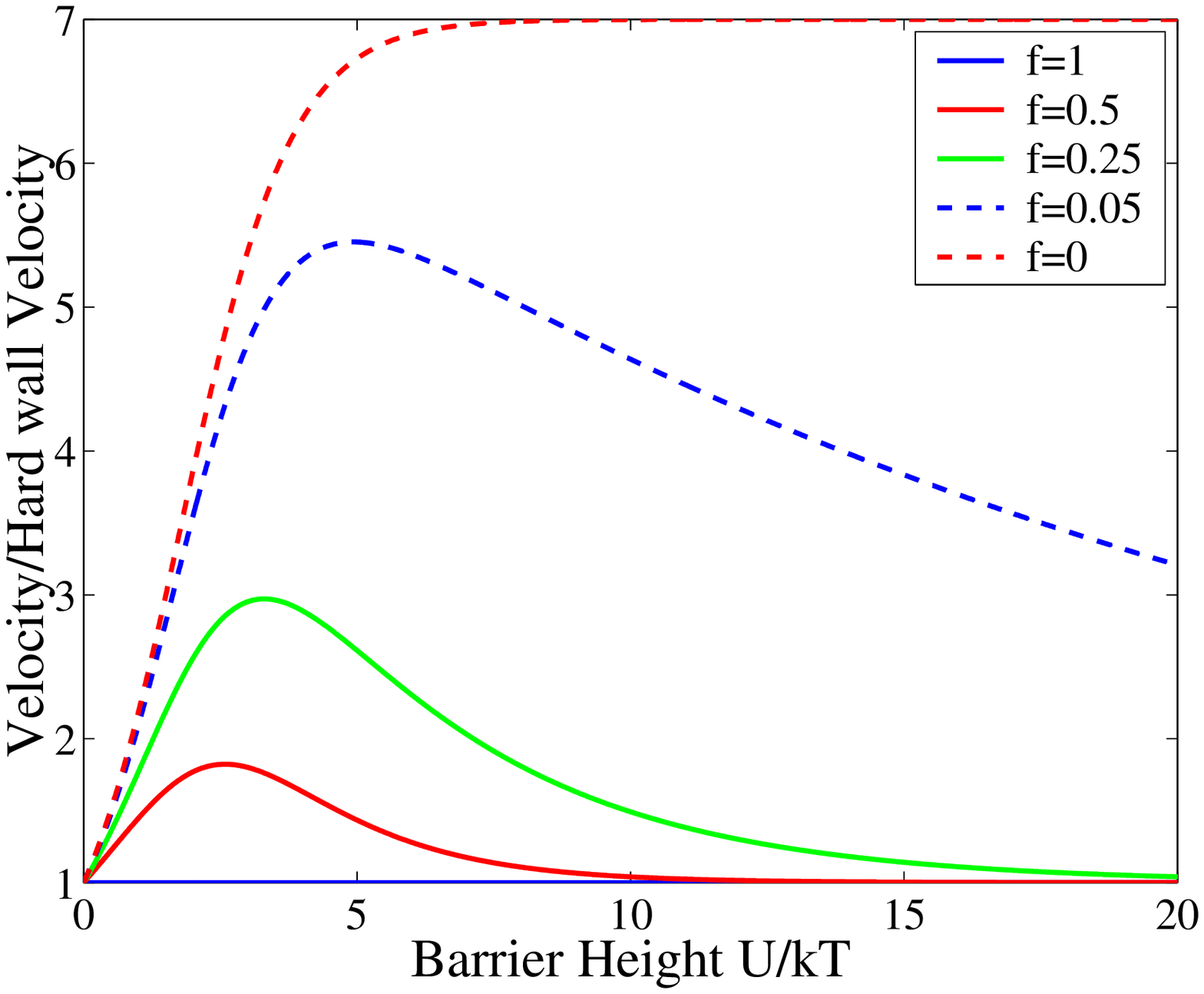}\includegraphics[height=5cm]{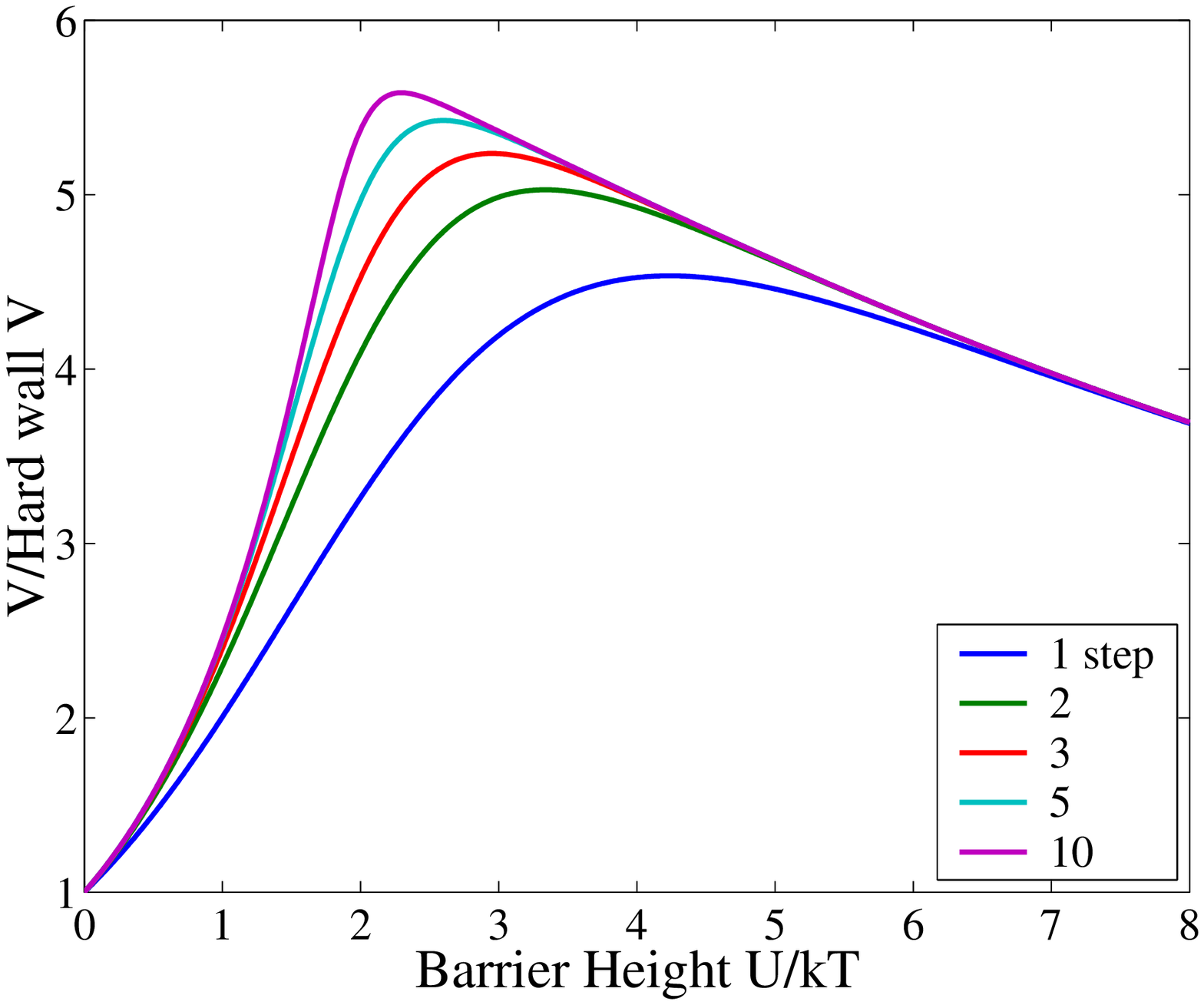}
\caption[]  
{Left, Ratio of the opening velocity with one-step potential to  
the hard-wall opening velocity versus height of the step. The  
different curves show different values of the weighting parameter  
$f$. The maximum opening speed increases as $f \to 0$, which  
corresponds to leaving the forward helicase and DNA closing rates  
unchanged, while increasing the helicase backward rate and the DNA  
opening rate.  Right, the  velocity  for $f =0.1$ versus height  
of the step, shown for staircases with different number of  
steps. }  
\label{onestepvel}  
\end{figure}

Next we consider a simple example of soft opening shown in Fig.  
\ref{enplot}b. Instead of a hard-wall potential, we insert a step of  
height $U_o$ at $j=0$ and a hard wall at $j=-1$. Thus the DNA junction  
and the helicase can overlap by one base if the energetic cost is  
paid. The ratios of the forward and backward rates at the step must  
satisfy the relations  
\begin{eqnarray}  
\frac{k^-_0}{k^+_{1}}&=&\frac{k ^-}{k ^ + } e^{U_0}, \\  
\frac{\alpha_0}{\beta_{1}}&=&\frac{\alpha}{\beta} e^{U_0}, \\  
k^+_0=\beta_0&=&0.  
\end{eqnarray}  
These relations affect only the ratios of the rates. We also have to 
specify how the rates themselves change. This is determined by  
the position of the barrier which must be crossed for the  
helicase to move or for the NA to open/close. We describe this by a  
weighting factor $0<f<1$.  (Note that $f$ is the same for the helicase  
motion and the NA breathing since both phenomena involve the same  
barrier.) Smaller values of $f$ imply that the barrier influences  
mainly backward rates, while for large $f$ the barrier affects forward  
rates.  We have  
\begin{eqnarray}  
k^+_{1}&=& k ^ +e^{- f U_o}, \\  
k^-_0&=&k ^-e^{-(f-1) U_o}, \\  
\beta_{1}&=&\beta e^{- f U_o},\\  
\alpha_0 &=& \alpha e^{-(f-1) U_o} .  
\end{eqnarray} 
With these values, we can calculate the opening velocity $v_1$ for one step. 
For comparison, we present the opening velocity relative to $v_{HW}$  
of Eq. (\ref{vhardwall}):  
\begin{equation}  
\frac{v_1}{v_{HW}}= \frac{c + (1-c)e^{-f U_o} }{c + (1-c) e^{- U_o} }.  
\label{onestep}  
\end{equation}  
where $c=\frac{\alpha+k^-}{\beta+k^+}$. Since $0<f<1$, the opening 
velocity $v_1$ for a single step is always larger than $v_{HW}$. 
However, the size of $v/v_{HW}$ depends strongly on the choice of $f$ 
(Fig. \ref{onestepvel}a). Smaller values of $f$ increase the opening 
velocity.  The maximal opening rate occurs when $f=0$. From 
Eq.(\ref{onestep}), we see that for $f=0$ and large $U_o$, $v_1 \to 
v_{HW}/c \approx 7 v_{HW}$.  For small values of $f$ the helicase 
forward rate $k ^ + $ and the DNA closing rate $\beta$ do not change 
much due to the step, whereas the helicase backward rate $k ^-$ and 
the DNA opening rate $\alpha$ increase significantly.  The main 
advantage of active opening is that it increases the opening rate of 
the DNA. It is more advantageous to increase the opening rate than to 
decrease the closing rate, because with an increased opening rate the 
strand-separation timescale is faster.

The unwinding rate also depends on the step height $U_o$. For small
$U_o$, the velocity is little changed from the hard-wall velocity. As
the step height increases, so does the opening velocity, as a result
of the increase in DNA opening rate caused by the presence of the
step. For a height of 1 kT and $f =0.05$, the opening velocity is
approximately twice the hard-wall velocity. We draw particular
attention to the barrier height of 2 kT. This means that the helicase
at $j=0$ causes the base-pair nearest the junction to be indifferent
to opening.  A 2 kT step height with $f=0.05$ increases the opening
velocity by approximately a factor of 3 relative to the hard-wall
velocity. For a further comparison of our calculation to experimental
data, please see the Discussion below.

\subsection{Multi-step staircase}  
  
Evidence from the crystal structures of PcrA bound to a forked DNA substrate  
discussed above suggests that we consider more than one ``step'' in the  
potential. The structures indicate that the helicase and dsDNA  
interact 4-5 bp from the junction, and again 12-13 bp from the  
junction. Thus  we should consider an interaction  
potential with a variable number of steps. For  
a multi-step staircase (with $n$ identical steps, each of height  
$U_o$, see Fig. 2c)  the unwinding velocity is  
  
\begin{equation}  
\frac{v_n}{v_{HW}}= \frac{c^{n} + (1-c) e^{-(f-1) U_o} \sum_{ j =1}^{ n}c^{ n-j}e^{-j U_o}}  
	{c^{n} + (1-c) \sum_{ j =1}^{ n}c^{ n-j}e^{-j U_o}}.  
\end{equation}  
This expression can be simplified by performing the  
summation, but as written one can see the similarity between this  
expression and Eq. (\ref{onestep}) for the one-step unwinding  
velocity. In particular, note that for $n=1$ the expression here  
reduces to Eq. (\ref{onestep}). Fig. \ref{onestepvel}b shows  
how the velocity changes with step number (for $f = 0.1 $). Adding  
additional steps allows the opening velocity to increase faster, and  
the maximum opening speed increases with the number of steps. We note  
that as the number of steps becomes large, the step size $U_o$  
corresponding the the maximum opening velocity approaches $ U_{max}  
\approx - \ln c$. In other words, in the limit of a large number of  
steps, the optimal step height approaches the free energy change of  
melting one base of NA: the fastest opening occurs when the  
interaction between helicase and NA causes the melting of one NA  
base pair to be {\em energetically neutral}.  
  
For a step height of 2 kT, adding one step increase the opening  
velocity by a factor of 3.5, whereas for 5 steps of height 2 kT the  
opening velocity is 5 times the hard-wall velocity. This is a
significant increase over the hard-wall velocity. The PcrA  
mutation studies \cite{sou00} found that mutating the residues which  
interact with the dsDNA decreased the unwinding rate by a factor of  
10-30. While our model shows a decrease in the opening velocity (when  
opening changes from soft to hard) of at most a factor of $1/c  
\approx 7$, we find it striking that such a simple model can  
demonstrate such a large change in the velocity. We emphasize that  
because our result looks at the ratio of active to passive unwinding  
rates, it is {\em independent} of particular helicase properties such  
as $k^+$ and $k^-$. Our result depends only on the sequence-averaged  
ratio of DNA opening and closing rates $c$ as well as the number of  
steps $n$ and the step size $U_o$.  
  
Indeed, while hard opening is significantly slower than the PcrA  
single-strand translocation rate, the helicase that chooses (or  
evolves) an appropriate interaction potential can open NA nearly as  
fast as the rate it moves on ssNA.  
  
\section{Discussion}  
\label{disc}  
  
In this paper, we have described the opening of NA by a helicase in  
terms of two hopping degrees of freedom: the position of the helicase,  
which hops along ssNA driven by ATP hydrolysis, and the position of the  
ss-ds NA junction, which fluctuates passively. This simplified  
model allows us to focus on how the motion is affected by the  
interaction between these two degrees of freedom.  We demonstrate that  
active opening in general leads to a faster overall unwinding rate  
that can approach the velocity of motion of helicase along the ssNA.  
  
We show that the interaction between helicase and dsNA determines the  
speed of unwinding. If rapid unwinding is advantageous, helicases will  
evolve towards interaction potentials which optimize this speed.  In  
our calculations we can make quantitative predictions of how different  
potentials change the unwinding rate.   
  
Because the helicase moves primarily forward, the backward rate $k^-$
is small compared to $k^+$.  Dillingham, Wigley, and Webb used two
different techniques\cite{dil00, dil02} to measure the rate of PcrA
motion on ssDNA. In our notation, this is $k ^ +-k ^-=80$
bases/sec. They found that their data were well-fit by a model with no
backward steps ($k ^-=0$), although our analysis of their results
found a comparably good fit with values of $k ^-$ up to 10\% of $k ^
+$.  The free-energy change $\Delta F$ of opening one base determines
the ratio $\alpha/\beta=e^{-\Delta F/kT}$.  This ratio is
approximately $\Delta F \simeq 2 kT $ for a base-pair near a
junction\cite{non95}. This value is consistent with bulk melting-curve
results
\cite{loh96} and is the origin of our number $\alpha/ \beta=  
e^{-2}\approx 1/7$. For the actual values of $\alpha$ and $\beta$, we
deduce a number using the experimental results of Bonnet, Krichevsky,
and Libchaber \cite{bon98} who found that the opening rate of a 5-bp
hairpin loop at 300 K and at 0.1 M NaCl was $k\simeq3000/$sec. Using a
simple kinetic model\footnote{We assume that each of the 5 bases has
an opening rate $\alpha$ and a closing rate $\beta$ and the ratio
$\alpha/\beta=c$. With $P_n$ denoting the relative probability of
having $n$ base pairs open, the opening rate is given by $k=J/\sum P_n
=\alpha c^{3}/(1 +2c+3c^2+4c^3+c^4(1+c+c^2+c^3+c^4))$, where $J=\alpha
P_0-\beta P_1=\alpha P_4$ is the current. With this expression we
estimate $\alpha$, given $k$ and $c$.}  we estimate from this value of
$k$ that $\alpha \approx 1.4 \times 10^6$/sec. We use this value of
$\alpha$ in our calculations below.  This estimate can be compared
with theoretical estimates of the base-pair opening rate $\alpha$,
which range from $10^5$/sec to $10^6$/sec
\cite{che92,kam97}.  
  
Here we compare the results of our model to experimental data, using  
the experimental values of $k^+$, $k^-$, $\alpha$, and $\beta$  
discussed above The maximum possible hard-wall opening velocity occurs  
in the case where $k^-=0$. Then  
\begin{equation}  
v_{HW}=\frac{\alpha}{\beta}\left(\frac{ k^+}{1+k^+/\beta}\right)\approx 11\   
\mbox{bases/sec}.  
\end{equation}  
Note that this is an upper bound on the velocity: any non-zero value  
of $k^-$ decreases the hard-wall velocity. Unfortunately, to our  
knowledge the unwinding rate has not been directly measured for SF1  
helicases such as PcrA. Existing unwinding assays \cite{sou00} use gel  
electrophoresis to determine the fraction of dsDNA molecules  
completely unwound at a given time. Therefore we cannot make a direct  
comparison between our calculation and measured unwinding velocities.  
  
Adding a single-step-staircase potential allows the velocity to
increase by up to a factor of 7. We choose a modest step height of 2
kT, which gives a maximum increase over the hard-wall velocity of a
factor of 3 (for $f=0.1$). Thus the DNA opening velocity could be up
to 33 bases/sec for a single-step-staircase potential.  For a
multi-step staircase, 5 steps of height 2 kT increase the opening rate
by a factor of 5. Thus the helicase unwinding rate can increase to 55
bases/sec from the hard rate of 11 bases/sec. This velocity is
nearly as large as the single-strand translocation speed itself. Thus
we demonstrate that soft opening---despite the extemely
simplified form of the interaction potential we have chosen---can
significantly increase the strand-separation rate.
  
We emphasize that changes to the helicase hopping rates $k^+$ and
$k^-$, for example due to changes in the ATP concentration, lead to
changes in the unwinding rate which our model predicts
quantitatively. Similarly, the free energy change of melting one NA
base pair can by changed by altering solution conditions or by
applying a force to the NA molecule. Such a change directly alters the
parameter $c \approx \alpha/\beta$ in our model, and we quantitatively
predict the resulting change in the unwinding rate.  Here, we have
neglected the effects of the base sequence on NA opening which are
believed to be weak for helicases\cite{loh96}. However, work by
Lubensky and Nelson \cite{lub00} suggest that interesting effects can
arise if a random DNA sequence is opened.

The simplified structure of this model means it may be useful in other
situations where two degrees of freedom interact.  For example, two
motor proteins which walk on a microtubule may affect each other's
motion. As mentioned above, our work may be viewed as a generalization
of the work of Peskin {\it et al.} \cite{pes93} to include an
arbitrary type of interaction potential. This generalization may also
be relevant to the problem originally addressed in Ref. \cite{pes93}:
the production of a force by a polymerizing filament. Introducing a
potential (which describes the interaction between the growing
filament tip and the obstacle against which the polymer pushes)
affects the polymerization speed and the force-velocity relation of
the filament.
  
We thank David Lubensky, Thomas Perkins, and Jacques Prost for helpful
discussions. MDB acknowledges the support of a Chateaubriand
Fellowship and funding from the VIGRE program of the NSF.
  
%Things for a longer paper:  
%\begin{enumerate}  
%\item four state model for helicase motion  
%\item Mention other applications of the model?  
%\item Sequence sensitivity -  ATP/ADP concentration  
%regime wherein the sequence sensitivity becomes important  
%\item Force-velocity curves  
%\item effective steps size of Bensimon  
%\item processivity  
%\end{enumerate}  
%  

\appendix  
\section{Calculation of the Unwinding Velocity}  
%mdb - better section title
\label{hopderive}  
Here we 
determine the rate of dsNA opening for the hopping model,  
given the rates $k^+$, $k^-$, $\alpha$, $\beta$, and the coupling  
potential $U(m-n)$. Consider the probability $P(n,m,t)$ that the  
helicase is at position $n$ and the ss-ds junction at position $m$ at  
time $t$. 
The probability satisfies
\begin{multline}  
\frac{\partial P(n,m)}{\partial t} = -(k^+_{m-n} +k^-_{m-n}  
	+\alpha_{m-n} +\beta_{m-n}) P(n,m)  +  \alpha_{m-n-1} P(n, m-1) \\  
	+ \beta_{m-n+1} P(n, m+1) + k^+_{m-n+1} P(n-1,m)  
	+k^-_{m-n-1} P(n+1,m).  
\label{basicprob}  
\end{multline}  
Note that the subscript $m-n$ means the hop starts at $m-n$ and  
moves either right or left.  
After rewriting using the difference and midpoint  
variables $j=m-n$ and $l=m+n$, we have
\begin{multline}  
\frac{\partial P(j,l)}{\partial t} = -(\alpha_{j} +\beta_{j}  
	+k^+_{j} +k^-_{j}) P(j,l) + \alpha_{j-1} P(j-1,l-1) +  
	\beta_{j+1} P(j+1,l+1) \\  
	+k^+_{j+1} P(j+1,l-1) +k^-_{j-1} P(j-1,l+1).  
\label{newprob}  
\end{multline} 
Since the coefficients in Eq. (\ref{newprob}) do not depend on $l$,
we can obtain an equation for the difference variable distribution
${\cal P}_j=\sum_j P(j,l)$ by summing over $l$
\begin{equation}  
\frac{\partial{\cal P}_j }{\partial t}  = -(k^+_{j} +k^-_{j}  
	+\alpha_{j} +\beta_{j}){\cal P}_j  
	+\alpha_{j-1} {\cal P}_{j-1}  
  	+ \beta_{j+1} {\cal P}_{j+1}+	k^+_{j+1}  
  	{\cal P}_{j+1}  +k^-_{j-1} {\cal P}_{j-1}.  
\label{jprob}  
\end{equation}  
This is  the equation for a position-dependent hopping model  
with forward rate $k^-_j+\alpha_j$ and backward rate  
$k^+_j+\beta_j$. The stationary probability distribution for $j$  
satisfies the recursion relation  %
$I(j)=I(j-1)$,
where we have defined the current  
\begin{equation}  
I(j)=(k^+_{j+1}+\beta_{j+1}) {\cal P}_{j+1} - (k^-_{j}+\alpha_{j})  
{\cal P}_{j}.  
\end{equation}  
The stationary probability distribution corresponds to a  
constant current solution. In the case of a potential $U(j) \to  
\infty$ as $j \to -\infty$, the current must be zero (no helicases can  
escape to $j \to -\infty$). This gives the zero-current relation  
\begin{equation}  
{\cal P}_{j+1} =  \frac{k^-_{j}+\alpha_{j}}{k^+_{j+1}+\beta_{j+1}} {\cal P}_{j}  
\label{nocurr}  
\end{equation}  
In the zero-current case, and if
the rates are are constant (independent of $j$), ${\cal P}_{j}$ has a
power law form with a constant determined by normalization.
For the general case where the rates $k^+_j$,  
$k^-_{j}$, $\alpha_{j}$, and $\beta_j$ vary with $j$, we can solve  
Eq. \ref{nocurr} iteratively, using the conditions  
(i) ${\cal P} \to 0$ as $\ j \to \infty$, and (ii) $\sum_j {\cal P}_j=1$.

The difference variable distribution relaxes to a steady state in a
finite time while the midpoint variable $l$ undergoes a drift with
diffusion. We can find a simple equation for the midpoint motion if we
focus on times longer than the relaxation time of ${\cal }P_j$.
(Since the DNA opening/closing times are of order of microseconds, we
expect the relaxation time to be at most hundreds of
microseconds. This time is short compared to the overall rate of
unwinding).
In this limit, the probability $P(j,l)$ becomes of the
form $P(j,l)={\cal P}_j \Pi_l$. 
If we define
\begin{eqnarray}  
p&=&\sum_j (\alpha_{j}+k^+_{j}){\cal P}_j,\\  
q&=&\sum_j (\beta_{j}+k^-_{j} ){\cal P}_j.  
\end{eqnarray}  
then the equation for $\Pi_l$ reduces to a hopping model with
effective forward and backward rates $p$ and $q$:
\begin{equation}  
\frac{\partial \Pi_l}{\partial t}=-(p+q) \Pi_l+p\ \Pi_{l-1}+q\ \Pi_{l+1}.  \label{lprob} 
\end{equation}  
The  rates $p$ and $q$ are {\em independent} of $l$, because of our  
assumption that all rates depend only on the difference variable  
$j$. As above, defining the current $\Upsilon(l)=p \Pi_l-q \Pi_l$  
the equation for the stationary distribution $\Pi$ becomes  
$\Upsilon(l)=\Upsilon(l-1).$
For the midpoint motion, there is no confining potential which  
localizes the probability distribution. Thus we expect a solution with  
non-zero current.
The solution is $\Pi_o= \frac{1}{N}$ and $\Upsilon=(p-q) \Pi_o$,
where $N$ is the total number of NA bases.  The mean opening velocity  
of a single helicase is  
\begin{equation}  
v =\frac{1}{2} (p-q) = \frac{1}{2}\sum_j (k^+_{j}  
	+\alpha_{j}-k^-_{j} -\beta_{j}){\cal P}_j.  
\label{appvel}  
\end{equation}  
As discussed in the text, $v$ is the unwinding rate at separation $j$ 
multiplied by the probability ${\cal P}_j$ of finding the complex at 
separation $j$. Repeating this addition for all possible $j$, we 
arrive at the total unwinding rate. The factor of 1/2 appears because 
we use $l=m+n$ for convenience, while the true midpoint location is 
$l/2$. Thus we see that, under our assumptions, solving Eq. 
\ref{nocurr} for the stationary probability distribution ${\cal 
P}_{j}$ immediately gives the {\em exact} mean helicase opening velocity. 
  
We can also determine the effective diffusion coefficient which  
characterizes velocity fluctuations  
\begin{equation}  
D=\frac{1}{4} \sum_j (k^+_{j} +\alpha_{j}+k^-_{j} +\beta_{j}){\cal P}_j  
\end{equation}  
  
%\bibliographystyle{pnas}  
%\bibliography{helicase}  

\end{document}